\documentclass[reprint, amsmath,amssymb, aps,pra]{revtex4-2}
\usepackage{stackengine}
\usepackage{booktabs}
\usepackage{float}

\usepackage[caption=false]{subfig}
\usepackage{graphicx}
\usepackage{bm}
\usepackage{hyperref}
\usepackage{natbib}
\usepackage[usenames,dvipsnames]{color}
\usepackage[usenames,dvipsnames,svgnames,table]{xcolor}
\usepackage{physics}
\usepackage{ulem}

\usepackage{comment}
\bibpunct{\color{blue}[}{\color{blue}]}{,}{n}{}{;}
\hypersetup{
  colorlinks,
  citecolor=blue,
  linkcolor=blue,
  urlcolor=blue
  }

\usepackage[T1]{fontenc}

\newlength{\mysize}

\begin{document}
\preprint{APS/123-QED}
\title{Reply to Antipov et al., Microsoft Quantum:
\\``Comment on Hess et al. Phys. Rev. Lett. 130, 207001 (2023)"}
\author{Henry F. Legg}
\author{Richard Hess}
\author{Daniel Loss}
\author{Jelena Klinovaja}%
\affiliation{%
 Department of Physics, University of Basel, Klingelbergstrasse 82, CH-4056 Basel, Switzerland }%
 \date{\today}
\begin{abstract} 
In this Reply we respond to the comment by Antipov {\it et al.}~\cite{Antipov2023} from Microsoft Quantum on Hess {\it et al.}, PRL 130, 207001 (2023)~\cite{Hess2023}. Antipov {\it et al.} reported only a single simulation and claimed it did not pass the Microsoft Quantum topological gap protocol (TGP). They have provided no parameters or data for this simulation (despite request). Regardless, in this reply we demonstrate that the trivial bulk gap reopening mechanism outlined in Hess {\it et al.}, in combination with trivial ZBPs, passes the TGP and therefore can result in TGP false positives. 
\end{abstract}

\maketitle

The concept of a topological gap protocol (TGP) that combines multiple expected features of Majorana bound states (MBSs) is appealing. However, the basis of the TGP set out by Microsoft Quantum is predicated on the claim that the bulk gap closing and reopening signal (BRS) in nonlocal conductance is a ``less ambiguous signature than the topological phase itself'' ~\cite{Pikulin2021}. Hess {\it et al.}~\cite{Hess2023},  demonstrated that a few overlapping Andreev bound states (ABSs) can mimic a topological BRS and that this can be combined with zero-bias peaks (ZBPs) to reproduce the main features of the TGP. In this reply we go further and show that the trivial BRS mechanism of Hess {\it et al.} can result in TGP false positives.

\begin{figure}[t!]
\includegraphics[width=1\columnwidth]{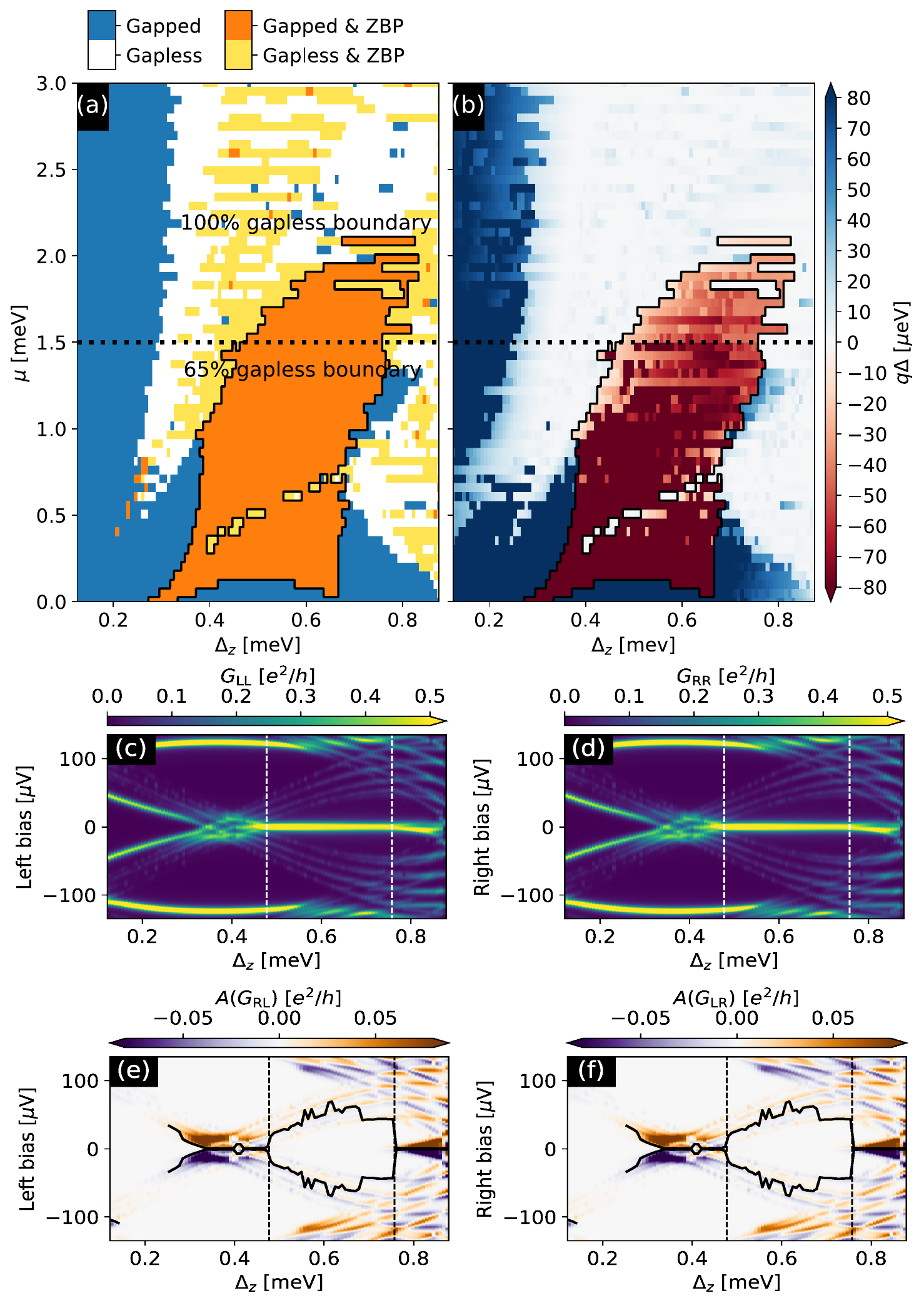}\vspace{-1pt}
\caption{{\bf TGP false positive:} (a-b) Result of Microsoft Quantum TGP~\cite{Github} with Hess {\it et al.} trivial BRS~\cite{Hess2023} and trivial ZBPs. Two false positive regions with 65\% and 100\% gapless boundaries are identified, see red regions in (b). (c-d) Local conductance with trivial ZBPs, along dashed line in (a-b). (e-f) Corresponding anti-symmetrised nonlocal conductance with Hess {\it et al.} trivial BRS due to a few overlapping ABSs. Parameters: $a=5~{\rm nm}; (N_L=N_R, N_S, N_N, N_{B, L}=N_{B, R})=(90,140,30,7), \; M=5; \; (t_L=t_R, t_{S N}, \mu_L=\mu_R, \mu_{S N}, \Delta_0, \Delta_Z^c, \alpha_L= \alpha_R, \gamma_L=\gamma_R, \mu_{\text {Lead}, L}=\mu_{\text {Lead}, R}) =(50,10,0,\mu,0.125,0.875,6.875,5,20)\mathrm{meV}; T=20~\mathrm{mK}$. $\Delta_{\rm max}=0.8 \Delta_{\rm Al}=200$~$\mu$eV~\cite{Pikulin2021}.}
\label{fig1}
\end{figure}

{\bf Trivial bulk gap reopening signature:} We first note that, in their comment, Antipov {\it et al.}~\cite{Antipov2023} do not dispute that the mechanism outlined in Hess {\it et al.} results in a trivial BRS, in fact, they purportedly replicate it.

{\bf TGP false positives:} In Fig.~\ref{fig1} we show that the combination of the Hess {\it et al.} trivial BRS mechanism with trivial ZBPs passes the TGP for a large portion of phase space. This demonstrates  that the trivial BRS of Hess {\it et~al.} can easily result in TGP false positives. 

\begin{figure}[t]
\includegraphics[width=1
\columnwidth]{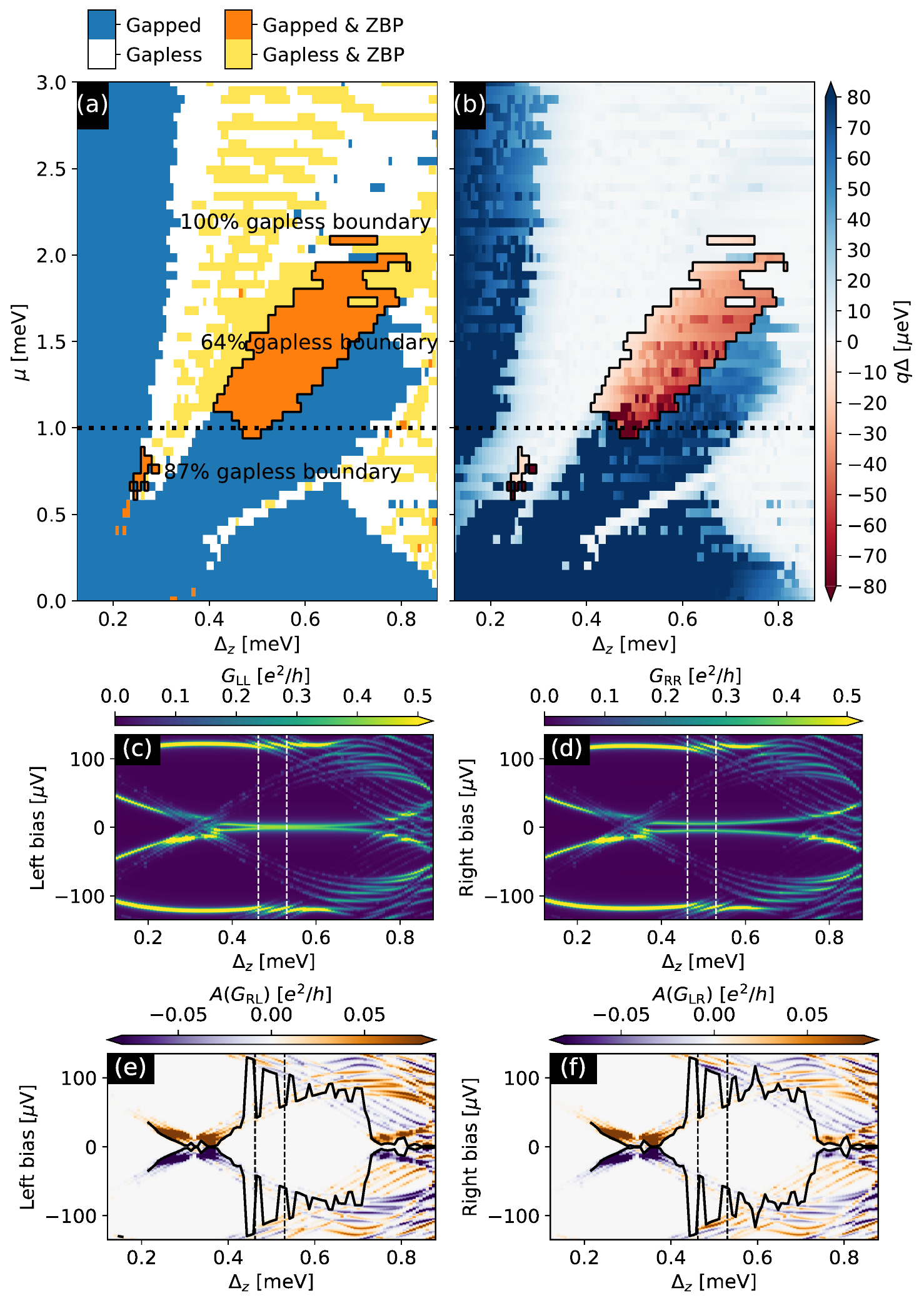}\vspace{0pt}
\caption{{\bf TGP false positive with split ZBP:} Example of the TGP applied to the Hess {\it et al.}~\cite{Hess2023} trivial BRS with a split ZBP. Three TGP false positive regions are indicated in (a-b).  Parameters: $a=5~{\rm nm}; (N_L,N_R, N_S, N_N, N_{B, L}=N_{B, R})=(90,94,140,30,7), \; M=5; \; (t_L=t_R, t_{S N}, \mu_L=\mu_R, \mu_{S N}, \Delta_0, \Delta_Z^c, \alpha_L, \alpha_R, \gamma_L=\gamma_R, \mu_{\text {Lead}, L}=\mu_{\text {Lead}, R}) =(50,10,0,\mu,0.125,0.875,6.875,6.6,5,20)\mathrm{meV}; T=10~\mathrm{mK}$, and $\Delta_{\rm max}=0.8 \Delta_{\rm Al}=200$~$\mu$eV~\cite{Pikulin2021}.}
\vspace{0pt}
\label{fig2}
\end{figure}

In their comment Antipov {\it et al.} report only a single simulation, for which they have provided no parameters or data (despite request). Furthermore, in their simulation Antipov {\it et al.} chose a ZBP that was split from zero-energy, the reason for this choice of split ZBP is unclear as it has no relevance to the trivial BRS of Hess {\it et al.} and Antipov {\it et al.} report that it is the primary reason their simulation does not pass the TGP. Moreover, even in combination with a split ZBP, the Hess {\it et al.} trivial BRS can still result in TGP false positives, see Fig.~\ref{fig2}.

{\bf TGP parameters:} It is important to note that the Microsoft Quantum TGP~\cite{Github} contains a large number of parameters. These range from parameters given in physical units, such as $\mu$eV, to parameters defined in number of data pixels. There are also considerable differences between the TGP of Pikulin {\it et al.}~\cite{Pikulin2021} compared to that utilised in Aghaee {\it et al.}~\cite{Aghaee2023}. In particular, the Aghaee {\it et al.} TGP is fine-tuned to the parameters of Microsoft Quantum's experiment and therefore is not generally applicable. We have, nonetheless, checked that the Hess {\it et al.} trivial BRS can result in TGP false positives using either the Pikulin {\it et al.} or Aghaee {\it et al.} TGP definition.

\newpage

\begin{figure}[h]
\includegraphics[width=1
\columnwidth]{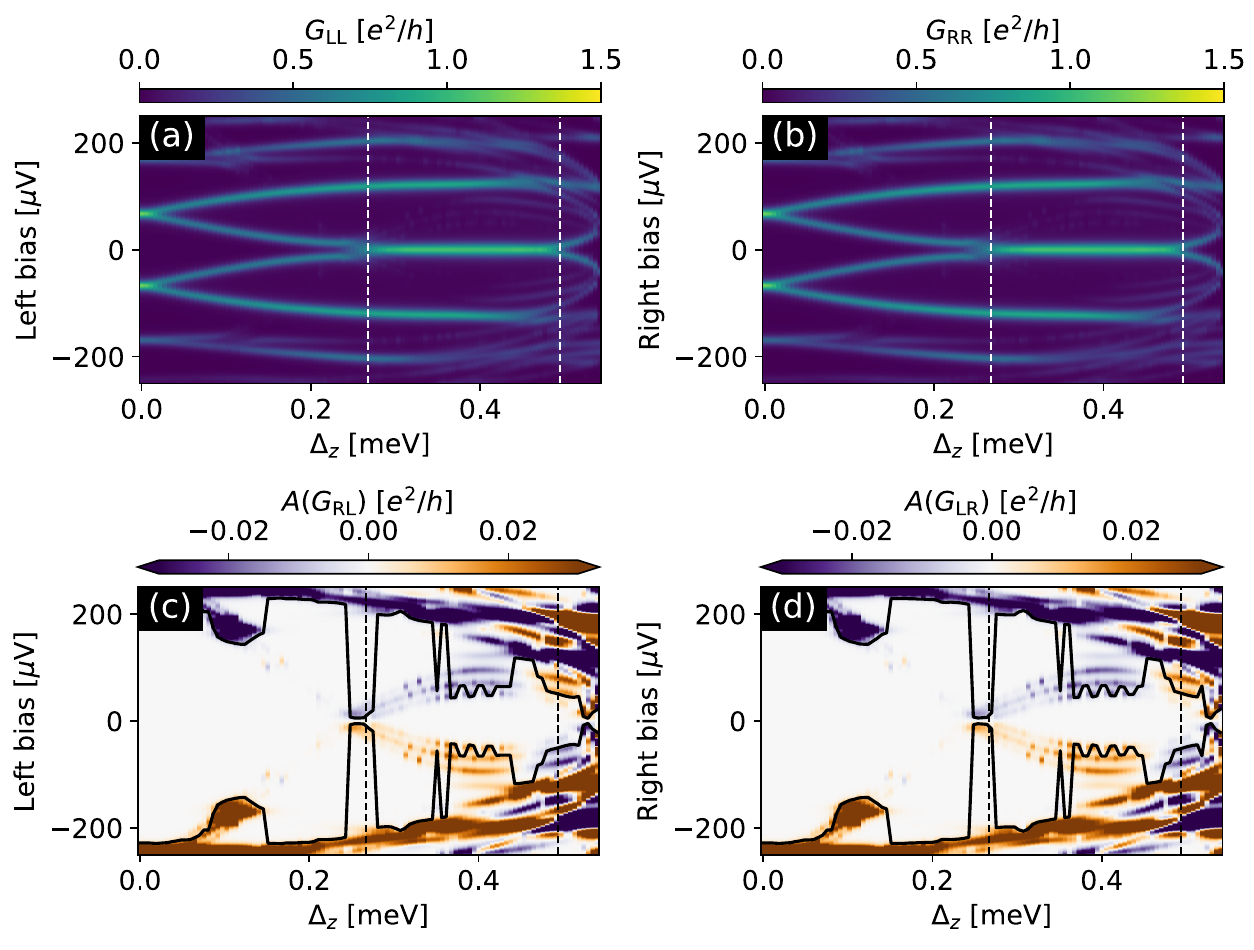}\vspace{-5pt}
\caption{{\bf Trivial BRS combined with trivial ZBPs due to quasi-MBSs:} This shows that the Hess {\it et al.}~\cite{Hess2023} trivial BRS can be combined with any trivial ZBP mechanism, here quasi-MBSs~\cite{Prada2012,Kells2012}, as long as it does not affect the formation of the Andreev band that results in the trivial BRS.}
\label{fig3}
\end{figure}

{\bf Trivial ZBPs:} A significant portion of the comment by Antipov {\it et al.} focuses on trivial ZBPs.  Since mechanisms for trivial ZBPs have been very extensively discussed, Hess {\it et al.}~\cite{Hess2023} made clear: ``the particular mechanism causing ZBPs at the ends of the nanowire is not the main subject of this paper and [our chosen] mechanism can be exchanged for any other that results in ZBPs, as long as the formation of the Andreev band is not affected''. To evince this statement, in Fig.~\ref{fig3}, we demonstrate that a different trivial ZBP mechanism (quasi-MBSs~\cite{Prada2012,Kells2012}) can be combined with the trivial BRS of Hess {\it et al.}

{\bf Relevance for realistic devices:} The claim by Antipov {\it et al.} that the Hess {\it et al.} trivial BRS mechanism requires ``a specific theoretical configuration of  Andreev states'' is false and ignores the extensive discussions of positional disorder in the main text and Supplemental Material ({\it e.g.} Fig.~S4) of Hess {\it et al.}~\cite{Hess2023}.

Antipov {\it et al.} further claim, without substantiation, that there is a ``vanishing probability'' for the Hess {\it et al.} trivial BRS to occur. However, current state-of-the-art devices contain many ABSs per coherence length and are only a few coherence lengths in size. As already discussed in Hess {\it et al.}, this means that a trivial BRS in current nanowires would only require two or three ABSs to overlap. Given the number of ABSs in current devices, this is not unlikely. Finally, we note that cross-hatch patterns occur in InAs devices grown on InP substrates~\cite{Lee2019} and can result in the quasi-periodic disorder that Hess {\it et al.} demonstrated can cause a trivial BRS in even extremely long nanowires.

{\bf Conclusion:} Antipov {\it et al.}~\cite{Antipov2023} do not dispute that the mechanism outlined in Hess {\it et al.}~\cite{Hess2023} results in a trivial BRS. In this reply we have demonstrated that the combination of the Hess {\it et al.} trivial BRS with trivial ZBPs results in TGP false positives. Given this demonstration and the lack of general applicability, we conclude that the Microsoft Quantum TGP utilised in Aghaee {\it et al.}~\cite{Aghaee2023} is neither sufficient nor necessary to identify a topological superconducting phase. 

\newpage
\bibliographystyle{apsrev4-2}
\bibliography{commentrefs}

\begin{thebibliography}{8}%
\makeatletter
\providecommand \@ifxundefined [1]{%
 \@ifx{#1\undefined}
}%
\providecommand \@ifnum [1]{%
 \ifnum #1\expandafter \@firstoftwo
 \else \expandafter \@secondoftwo
 \fi
}%
\providecommand \@ifx [1]{%
 \ifx #1\expandafter \@firstoftwo
 \else \expandafter \@secondoftwo
 \fi
}%
\providecommand \natexlab [1]{#1}%
\providecommand \enquote  [1]{``#1''}%
\providecommand \bibnamefont  [1]{#1}%
\providecommand \bibfnamefont [1]{#1}%
\providecommand \citenamefont [1]{#1}%
\providecommand \href@noop [0]{\@secondoftwo}%
\providecommand \href [0]{\begingroup \@sanitize@url \@href}%
\providecommand \@href[1]{\@@startlink{#1}\@@href}%
\providecommand \@@href[1]{\endgroup#1\@@endlink}%
\providecommand \@sanitize@url [0]{\catcode `\\12\catcode `\$12\catcode
  `\&12\catcode `\#12\catcode `\^12\catcode `\_12\catcode `\%12\relax}%
\providecommand \@@startlink[1]{}%
\providecommand \@@endlink[0]{}%
\providecommand \url  [0]{\begingroup\@sanitize@url \@url }%
\providecommand \@url [1]{\endgroup\@href {#1}{\urlprefix }}%
\providecommand \urlprefix  [0]{URL }%
\providecommand \Eprint [0]{\href }%
\providecommand \doibase [0]{https://doi.org/}%
\providecommand \selectlanguage [0]{\@gobble}%
\providecommand \bibinfo  [0]{\@secondoftwo}%
\providecommand \bibfield  [0]{\@secondoftwo}%
\providecommand \translation [1]{[#1]}%
\providecommand \BibitemOpen [0]{}%
\providecommand \bibitemStop [0]{}%
\providecommand \bibitemNoStop [0]{.\EOS\space}%
\providecommand \EOS [0]{\spacefactor3000\relax}%
\providecommand \BibitemShut  [1]{\csname bibitem#1\endcsname}%
\let\auto@bib@innerbib\@empty
\bibitem [{\citenamefont {Antipov}\ \emph {et~al.}(2023)\citenamefont
  {Antipov}, \citenamefont {Cole}, \citenamefont {Kalashnikov}, \citenamefont
  {Karimi}, \citenamefont {Lutchyn}, \citenamefont {Nayak}, \citenamefont
  {Pikulin},\ and\ \citenamefont {Winkler}}]{Antipov2023}%
  \BibitemOpen
  \bibfield  {author} {\bibinfo {author} {\bibfnamefont {A.}~\bibnamefont
  {Antipov}}, \bibinfo {author} {\bibfnamefont {W.}~\bibnamefont {Cole}},
  \bibinfo {author} {\bibfnamefont {K.}~\bibnamefont {Kalashnikov}}, \bibinfo
  {author} {\bibfnamefont {F.}~\bibnamefont {Karimi}}, \bibinfo {author}
  {\bibfnamefont {R.}~\bibnamefont {Lutchyn}}, \bibinfo {author} {\bibfnamefont
  {C.}~\bibnamefont {Nayak}}, \bibinfo {author} {\bibfnamefont
  {D.}~\bibnamefont {Pikulin}},\ and\ \bibinfo {author} {\bibfnamefont
  {G.}~\bibnamefont {Winkler}},\ }\href@noop {} {\  (\bibinfo {year} {2023})},\
  \Eprint {https://arxiv.org/abs/2307.15813} {arXiv:2307.15813} \BibitemShut
  {NoStop}%
\bibitem [{\citenamefont {Hess}\ \emph {et~al.}(2023)\citenamefont {Hess},
  \citenamefont {Legg}, \citenamefont {Loss},\ and\ \citenamefont
  {Klinovaja}}]{Hess2023}%
  \BibitemOpen
  \bibfield  {author} {\bibinfo {author} {\bibfnamefont {R.}~\bibnamefont
  {Hess}}, \bibinfo {author} {\bibfnamefont {H.~F.}\ \bibnamefont {Legg}},
  \bibinfo {author} {\bibfnamefont {D.}~\bibnamefont {Loss}},\ and\ \bibinfo
  {author} {\bibfnamefont {J.}~\bibnamefont {Klinovaja}},\ }\href
  {https://link.aps.org/doi/10.1103/PhysRevLett.130.207001} {\bibfield
  {journal} {\bibinfo  {journal} {Phys. Rev. Lett.}\ }\textbf {\bibinfo
  {volume} {130}},\ \bibinfo {pages} {207001} (\bibinfo {year}
  {2023})}\BibitemShut {NoStop}%
\bibitem [{\citenamefont {Pikulin~{\it et al.}}(2021)}]{Pikulin2021}%
  \BibitemOpen
  \bibfield  {author} {\bibinfo {author} {\bibfnamefont {D.~I.}\ \bibnamefont
  {Pikulin~{\it et al.}}},\ }\href@noop {} {\  (\bibinfo {year} {2021})},\
  \Eprint {https://arxiv.org/abs/2103.12217} {arXiv:2103.12217} \BibitemShut
  {NoStop}%
\bibitem [{\citenamefont {{Microsoft Quantum TGP as downloaded on
  02.08.23}}()}]{Github}%
  \BibitemOpen
  \bibfield  {author} {\bibinfo {author} {\bibnamefont {{Microsoft Quantum TGP
  as downloaded on 02.08.23}}},\ }\href
  {https://github.com/microsoft/azure-quantum-tgp} {\bibinfo  {journal}
  {{github.com/microsoft/azure-quantum-tgp}\hspace{-3pt}}\ }\BibitemShut
  {NoStop}%
\bibitem [{\citenamefont {Aghaee~{\it et al.}}(2023)}]{Aghaee2023}%
  \BibitemOpen
\bibfield  {journal} {  }\bibfield  {author} {\bibinfo {author} {\bibfnamefont
  {M.}~\bibnamefont {Aghaee~{\it et al.}}} (\bibinfo {collaboration} {Microsoft
  Quantum}),\ }\href {https://doi.org/10.1103/PhysRevB.107.245423} {\bibfield
  {journal} {\bibinfo  {journal} {Phys. Rev. B}\ }\textbf {\bibinfo {volume}
  {107}},\ \bibinfo {pages} {245423} (\bibinfo {year} {2023})}\BibitemShut
  {NoStop}%
\bibitem [{\citenamefont {Prada}\ \emph {et~al.}(2012)\citenamefont {Prada},
  \citenamefont {San-Jose},\ and\ \citenamefont {Aguado}}]{Prada2012}%
  \BibitemOpen
  \bibfield  {author} {\bibinfo {author} {\bibfnamefont {E.}~\bibnamefont
  {Prada}}, \bibinfo {author} {\bibfnamefont {P.}~\bibnamefont {San-Jose}},\
  and\ \bibinfo {author} {\bibfnamefont {R.}~\bibnamefont {Aguado}},\ }\href
  {https://doi.org/10.1103/PhysRevB.86.180503} {\bibfield  {journal} {\bibinfo
  {journal} {Phys. Rev. B}\ }\textbf {\bibinfo {volume} {86}},\ \bibinfo
  {pages} {180503} (\bibinfo {year} {2012})}\BibitemShut {NoStop}%
\bibitem [{\citenamefont {Kells}\ \emph {et~al.}(2012)\citenamefont {Kells},
  \citenamefont {Meidan},\ and\ \citenamefont {Brouwer}}]{Kells2012}%
  \BibitemOpen
  \bibfield  {author} {\bibinfo {author} {\bibfnamefont {G.}~\bibnamefont
  {Kells}}, \bibinfo {author} {\bibfnamefont {D.}~\bibnamefont {Meidan}},\ and\
  \bibinfo {author} {\bibfnamefont {P.~W.}\ \bibnamefont {Brouwer}},\ }\href
  {https://doi.org/10.1103/PhysRevB.86.100503} {\bibfield  {journal} {\bibinfo
  {journal} {Phys. Rev. B}\ }\textbf {\bibinfo {volume} {86}},\ \bibinfo
  {pages} {100503} (\bibinfo {year} {2012})}\BibitemShut {NoStop}%
\bibitem [{\citenamefont {Lee}\ \emph {et~al.}(2019)\citenamefont {Lee},
  \citenamefont {Shojaei}, \citenamefont {Pendharkar}, \citenamefont {Feldman},
  \citenamefont {Mukherjee},\ and\ \citenamefont {Palmstr\o{}m}}]{Lee2019}%
  \BibitemOpen
  \bibfield  {author} {\bibinfo {author} {\bibfnamefont {J.~S.}\ \bibnamefont
  {Lee}}, \bibinfo {author} {\bibfnamefont {B.}~\bibnamefont {Shojaei}},
  \bibinfo {author} {\bibfnamefont {M.}~\bibnamefont {Pendharkar}}, \bibinfo
  {author} {\bibfnamefont {M.}~\bibnamefont {Feldman}}, \bibinfo {author}
  {\bibfnamefont {K.}~\bibnamefont {Mukherjee}},\ and\ \bibinfo {author}
  {\bibfnamefont {C.~J.}\ \bibnamefont {Palmstr\o{}m}},\ }\href
  {https://link.aps.org/doi/10.1103/PhysRevMaterials.3.014603} {\bibfield
  {journal} {\bibinfo  {journal} {Phys. Rev. Mater.}\ }\textbf {\bibinfo
  {volume} {3}},\ \bibinfo {pages} {014603} (\bibinfo {year}
  {2019})}\BibitemShut {NoStop}%
\end{thebibliography}%


\begin{thebibliography}{1}%
\makeatletter
\providecommand \@ifxundefined [1]{%
 \@ifx{#1\undefined}
}%
\providecommand \@ifnum [1]{%
 \ifnum #1\expandafter \@firstoftwo
 \else \expandafter \@secondoftwo
 \fi
}%
\providecommand \@ifx [1]{%
 \ifx #1\expandafter \@firstoftwo
 \else \expandafter \@secondoftwo
 \fi
}%
\providecommand \natexlab [1]{#1}%
\providecommand \enquote  [1]{``#1''}%
\providecommand \bibnamefont  [1]{#1}%
\providecommand \bibfnamefont [1]{#1}%
\providecommand \citenamefont [1]{#1}%
\providecommand \href@noop [0]{\@secondoftwo}%
\providecommand \href [0]{\begingroup \@sanitize@url \@href}%
\providecommand \@href[1]{\@@startlink{#1}\@@href}%
\providecommand \@@href[1]{\endgroup#1\@@endlink}%
\providecommand \@sanitize@url [0]{\catcode `\\12\catcode `\$12\catcode
  `\&12\catcode `\#12\catcode `\^12\catcode `\_12\catcode `\%12\relax}%
\providecommand \@@startlink[1]{}%
\providecommand \@@endlink[0]{}%
\providecommand \url  [0]{\begingroup\@sanitize@url \@url }%
\providecommand \@url [1]{\endgroup\@href {#1}{\urlprefix }}%
\providecommand \urlprefix  [0]{URL }%
\providecommand \Eprint [0]{\href }%
\providecommand \doibase [0]{https://doi.org/}%
\providecommand \selectlanguage [0]{\@gobble}%
\providecommand \bibinfo  [0]{\@secondoftwo}%
\providecommand \bibfield  [0]{\@secondoftwo}%
\providecommand \translation [1]{[#1]}%
\providecommand \BibitemOpen [0]{}%
\providecommand \bibitemStop [0]{}%
\providecommand \bibitemNoStop [0]{.\EOS\space}%
\providecommand \EOS [0]{\spacefactor3000\relax}%
\providecommand \BibitemShut  [1]{\csname bibitem#1\endcsname}%
\let\auto@bib@innerbib\@empty
\bibitem [{Note1()}]{Note1}%
  \BibitemOpen
  \bibinfo {note} {\lowercase {h}ttps://github.com/microsoft/azure-quantum-tgp
  Timestamp: 17.09.23}\BibitemShut {NoStop}%
\end{thebibliography}%


%
\end{document}